\documentstyle[12pt]{article}
\begin{document}
\begin{titlepage}
\begin{flushright}   NSF-ITP-93-91;
                     gr-qc/9307002\\
\end{flushright}
\begin{center}
   \vskip 3em
  {\LARGE Conformal Invariance of Black Hole Temperature }
  \vskip 1.5em
  {\large Ted Jacobson\footnote{jacobson@umdhep} and
   Gungwon Kang\footnote{eunjoo@wam.umd.edu} \\[.5em]}
\em{ Institute for Theoretical Physics, University of California,
Santa Barbara, CA 93106\\
Department of Physics, University of Maryland, College Park,
MD 20742-4111}\\[.7em]
\end{center}
\vskip 1em
\begin{abstract}
It is shown that the surface gravity and temperature of a stationary
black hole are invariant under conformal transformations of the metric
that are the identity at infinity. More precisely, we find a conformal
invariant definition of the surface gravity of a conformal Killing
horizon that agrees with the usual definition(s) for a true Killing
horizon and is proportional to the temperature as defined by Hawking
radiation. This result is reconciled with the intimate relation between
the trace anomaly and the Hawking effect, despite the {\it non}invariance
of the trace anomaly under conformal transformations.

\end{abstract}
\end{titlepage}
\newpage
\def\L{{\cal L}}
\def\wt{\widetilde}
\section{Introduction}

Under a conformal transformation $g_{ab}\rightarrow\Omega^2 g_{ab}$
a black hole spacetime remains a black hole with the same event
horizon, at least if the conformal factor $\Omega^2$ is regular on the
event horizon and goes to unity at null infinity. This is simply because
the causal structure is invariant under conformal transformation.
Thus it makes sense to study the effect of conformal transformations
on the thermodynamical properties of black holes.

However, since the Einstein equation is not conformally invariant, a
conformally transformed Einstein black hole solution will not be a
solution to the field equations. Nevertheless, the transformed black
hole may still serve as a background on which Hawking radiation may
occur. Moreover, in gravitation theories (such as Brans-Dicke theory
or string theory) which contain a dilaton field $\phi$, both the
``Einstein metric" $g_{ab}$ and the metric $e^{2\phi}g_{ab}$ are
relevant for different considerations, and one may wish to know how
the thermodynamic quantities associated with these two metrics are
related.

In particular, we shall focus on the temperature and surface gravity
$\kappa$ of a black hole. The surface gravity plays the role of
temperature in classical black hole thermodynamics, and it is
proportional to the Hawking temperature $T_H=\kappa/2\pi$ which
characterizes the radiation emitted by a black hole via quantum
particle production. It is trivial to see that the Hawking radiation
of a free, conformally coupled field is invariant under conformal
transformations that are the identity at infinity. The form of the
radiation can be computed\cite{HawkRad} by evaluating the Bogoliubov
coefficients that relate the ingoing to the outgoing positive and
negative frequency mode functions in the black hole spacetime. This
computation involves only the classical propagation of the field, so
if the field is conformally coupled, the result is invariant under
conformal transformations.

It follows that in those cases where the surface gravity is defined,
it too must be conformally invariant. It is not so obvious why this is
so however, when phrased simply as a property of surface gravity.
Consider for example an asymptotically flat, static spacetime with
Killing field $\xi^a$, whose norm $V=(-\xi^a\xi_a)^{1/2}$ goes to
unity at infinity. The force that must be exerted at infinity in order
to hold a unit mass test particle fixed on an orbit of $\xi^a$ is just
$g=(\nabla^a V \nabla_a V)^{1/2}$, evaluated on that orbit. This is
the ``surface gravity" at the position of the orbit. Under a conformal
transformation $g_{ab}\rightarrow \Omega^2 g_{ab}$, one has
$g\rightarrow
[\Omega^{-2}g^{ab}\nabla_a(\Omega V) \nabla_b (\Omega V)]^{1/2}$,
so $g$ is clearly {\it not} conformal invariant! However,
we shall see below that if the conformal transformation is static,
then the surface gravity {\it at the horizon} is indeed conformal
invariant.\footnote{This is not quite as obvious as it may appear,
since $\nabla_aV$ is singular at the horizon where $V\rightarrow 0$.
In fact, $2V\nabla_a V=\nabla_a V^2$ is non-vanishing but tangent
to $\xi^a$ at the horizon. Thus if $\xi^a\nabla_a\Omega=0$ at the
horizon, $g$ will be invariant there.}

The definition of surface gravity in terms of acceleration of static
particles can be generalized to stationary spacetimes, using zero
angular momentum particles. One can also define surface gravity of a
Killing horizon directly in terms of the Killing field that is null on
the horizon (which amounts to the same thing), or by reference to the
periodicity of the analytically continued Killing time coordinate that
is required in order that the Euclidean section of the black hole
spacetime be non-singular.

All of these definitions of surface gravity are meaningful, and all
agree, for a stationary black hole whose event horizon is a {\it
Killing horizon}. A Killing horizon is a null hypersurface whose null
geodesic generators are orbits of a Killing field. A theorem of
Hawking\cite{HE} shows that in four dimensional Einstein gravity, a
stationary black hole event horizon is always a Killing horizon. As
far as we are aware, this result has not been generalized to spacetime
dimensions other than four or to theories other than Einstein's. (The
physical idea behind it\cite{Kip} is that if the generators are {\it
not} Killing orbits, the horizon will be bumpy and will radiate
gravitational waves, violating the assumption of stationarity.) Of
course it is certainly not true if field equations are not imposed;
for instance, it is violated if one subjects a stationary Einstein
black hole with a Killing vector $\chi^a$ that is null at the horizon
to a stationary conformal transformation with $\L_\chi\Omega\ne 0$.

Indeed, suppose $\chi^a$ is a Killing vector of the metric $g_{ab}$,
so that $\L_{\chi}g_{ab}=0$. Then for a conformally related metric
$\wt{g}_{ab}=\Omega^2 g_{ab}$, one has
\begin{displaymath}
\L_{\chi}\wt{g}_{ab}=(\L_{\chi}\Omega^2)g_{ab}
=(\L_{\chi}\ln\Omega^2)\wt{g}_{ab}.
\end{displaymath}
If the conformal factor is constant along the Killing orbits,
$\L_{\chi}\Omega=0$, then $\chi^a$ is a Killing field for
$\wt{g}_{ab}$ as well. Otherwise, $\chi^a$ is only a
{\it conformal Killing field}, and a Killing horizon is tranformed
into what we shall call a {\it conformal Killing horizon}.

Conversely, it is easy to see that a spacetime with a conformal
Killing field $\chi^a$ is conformal to a spacetime for which $\chi^a$
is a true  Killing field. That is, if
$\L_{\chi}\wt{g}_{ab}=2f\, \wt{g}_{ab}$,
then the transformed metric $g_{ab}=\Omega^2\wt{g}_{ab}$
will satisfy $\L_{\chi}g_{ab}=0$ provided $\Omega$ is chosen to be a
solution to the equation $\L_{\chi}\Omega^2 + 2f \Omega^2=0$. The
solutions are given along integral curves of $\chi^a$ by
$\ln\Omega=-\int f dv$, where $\chi^a \nabla_a v=1$.

In this letter we shall show that there is a {\it conformal-invariant}
definition $\kappa_1$ of the surface gravity of a conformal Killing
horizon that agrees with all the usual definitions of surface gravity
in the case of the Killing horizon of a stationary black hole. Since
it is conformal-invariant, $\kappa_1$ is identical to the surface
gravity of a conformally related true Killing horizon. Thus it is
clear that $\kappa_1/2\pi$ gives the correct Hawking temperature for
radiation emitted by a the conformal transform of a stationary black
hole.

As a final introductory remark, note that the area of the black hole
horizon is {\it not} invariant under conformal transformation. This
means that for example with the entropy given by one quarter the
surface area, the first and second laws are not invariant. Of course
we don't expect that they should be invariant, since the dynamical
equations of the theory are not conformal invariant.

\section{Surface gravity}

We would like to define the surface gravity of a conformal Killing
horizon in some interesting and useful way. The first question that
arises is whether the surface gravity should be thought of as a
property of the conformal Killing horizon itself, or whether a
particular conformal Killing field must be selected before the concept
of surface gravity even becomes well defined. A simple example
demonstrates that in fact a particular conformal Killing field must be
specified. The example is two dimensional Minkowski spacetime. Any
null line is a Killing horizon with respect to both a null translation
Killing field and a boost Killing field. Both of these Killing fields
can be used to define a ``surface gravity" for the horizon. With the
translation, the surface gravity vanishes, whereas with the boost one
can obtain any positive value depending on the overall scale of the
Killing vector. Thus the surface gravity should not be meaningful
until a particular Killing field is selected. In an asymptotically
flat spacetime one can select a Killing field, or perhaps a conformal
Killing field, by a boundary condition at infinity.

Suppose that a conformal Killing vector $\chi^a$ is null on some
conformal Killing horizon. Then, at the horizon, we can consider the
following three candidate definitions of surface gravity:

\begin{equation}
\nabla_a(\chi^b \chi_b)=-2\kappa_1 \chi_a
\label{k1}
\end{equation}

\begin{equation}
\chi^b \nabla_b \chi^a=\kappa_2 \chi^a
\label{k2}
\end{equation}

\begin{equation}
(\kappa_3)^2=-\frac{1}{2}(\nabla^a\chi^b)(\nabla_{[a} \chi_{b]})
\label{k3}
\end{equation}

The first quantity, $\kappa_1$, is well-defined since $\chi^2=0$
everywhere on the horizon, so its gradient must be proportional to the
normal to the horizon, which is $\chi_a$ itself. The second quantity,
$\kappa_2$, is well-defined since the horizon is a null hypersurface
whose null generators are therefore\cite{Wald} geodesics. The third
quantity, $\kappa_3$, is obviously well-defined, but what needs
explanation is the antisymmetrization of the $ab$ index pair. When
$\chi^a$ is a true Killing vector, $\nabla_a\chi_b$ is already
antisymmetric, and can be thought of as the infinitesimal generator of
an isometry (Lorentz tranformation) in the tangent space. In the case
of a Killing horizon this isometry is a boost. For a conformal Killing
vector, one has
\begin{equation}
2\nabla_{(a}\chi_{b)}= \L_\chi g_{ab}= 2f\, g_{ab}.
\label{f}
\end{equation}
This symmetric part can be thought of as the infinitesimal generator of a
dilatation in the tangent space. If one wants a definition that captures
only the quantity related to the local acceleration along the conformal
Killing flow, it makes sense to discard this symmetric part.

It is easy to determine the relationship between these three quantities.
In terms of the function $f$ defined in (\ref{f}) above, one has
\begin{equation}
\kappa_1=\kappa_2-2f=\kappa_3-f.
\label{relations}
\end{equation}
(In relating $\kappa_3$ to the others one uses the fact that
$\chi_{[a}\nabla_b \chi_{c]}=0$ at the horizon, which holds because
$\chi^a$ is orthogonal to a hypersurface (the horizon) there.)
For a true Killing field, $f$ vanishes, and all definitions agree.

Now let us consider the effect of making a conformal transformation.
Of course $\chi^a$ remains a conformal Killing field and the conformal
Killing horizon remains such. We wish to determine how the three
quantities $\kappa_i$ transform, assuming that they are computed with
respect to the {\it original} conformal Killing vector $\chi^a$. (In
the asymptotically flat case, the original conformal Killing field can
be determined by its ``initial data" \cite{Geroch} at infinity, which
can be considered fixed if the conformal factor goes to unity at
infinity.)

Since in general $f$ changes under a conformal transformation, at most
one of the quantities $\kappa_i$ can be conformally invariant. In fact,
$\kappa_1$ is the winner:
\begin{eqnarray}
\wt{\nabla}_a(\wt{g}_{bc}\chi^b\chi^c)
&=&\nabla_a(\Omega^2 g_{bc} \chi^b\chi^c)\cr
&=&-2\Omega^2 \kappa_1 g_{ab} \chi^b
+(\nabla_a\Omega^2) g_{bc} \chi^b\chi^c\cr
&=&-2\kappa_1 \wt{g}_{ab}\chi^b,
\end{eqnarray}
so that $\wt{\kappa}_1=\kappa_1$, provided that $\nabla_a\Omega^2$
is nonsingular at the horizon.

Conformal invariance of $\kappa_1$ implies via (\ref{relations})
invariance of $\kappa_2$ and $\kappa_3$ under those a conformal
transformations that are constant along the Killing field, since
$\L_\chi \Omega=0$ implies that the function $f$ in (\ref{f}) is
unchanged.

Two more definitions of black hole surface gravity are referred to
commonly, which are equivalent to $\kappa_{1,2,3}$ for the case of
stationary black holes with Killing horizons. The force per unit mass
$\kappa_4$ that must be applied at infinity to hold a zero angular
momentum particle ``at rest" just outside the horizon is one of these.
The other is $\kappa_5=2\pi/\beta$, where $\beta$ is the period of the
imaginary time coordinate required by regularity at the horizon of the
Euclidean section obtained by analytic continuation\cite{GibbPerr}.
Both of these definitions make sense only when the spacetime is
stationary.

Under stationary conformal transformations that go to unity at
infinity, $\kappa_5$ is clearly invariant, since the regularity
condition imposed at a point of the Euclidean horizon merely states
that circumference of an infinitesimal circle with center on the
horizon is $2\pi$ times its radius, and both of these simply scale
with the value of the conformal factor at the center of the circle.

Conformal invariance of the force per unit mass definition $\kappa_4$
is not quite so obvious. Indeed, as mentioned in the introduction, it
is {\it not} conformal invariant, except in the limit where the test
particle approaches the horizon. In the static case, the test particle
follows an orbit of the Killing field $\chi^a$, so has velocity
$u^a=(-\chi^d\chi_d)^{-1/2}\chi^a$ and acceleration $a^c=u^b\nabla_b u^c$.
Thus one has
$\kappa_4^2= {\rm lim}\{-\chi^a\chi_a)(a^c a_c)\}
={\rm lim}
\{(\chi^b\nabla_b\chi^c) (\chi^a\nabla_a\chi_c)/(-\chi^d\chi_d)\}$
as the horizon is approached.
As shown for example in Ref.~\cite{Wald}, this expression is equal to
$\kappa_3^2$. As shown above, $\kappa_3$ is invariant under a static
conformal transformation.

More generally in the stationary but nonstatic case, one has a time
translation Killing field $\xi^a$ and a rotation Killing field
$\psi^a$. Let the constant $\omega_r$ be chosen so that
$\zeta^a=\xi^a +\omega_r \psi^a$
satisfies $\zeta^a\psi_a=0$ at some radius $r$, so
the integral curve of $\zeta^a$ at $r$ corresponds to that of a zero
angular momentum test particle. One can show that the force per unit
mass that must be applied at infinity to hold such a test particle on
this worldline is given by
$F_\infty=\{(-\zeta^a\zeta_a)(a^c a_c)\}^{1/2}$,
where $a^c$ is the acceleration of the worldline. In
the limit as the horizon is approached, $\zeta^a$ approaches the
Killing field $\chi^a$ that is null on the horizon, and one has
$\kappa_4^2\equiv{\rm lim}\, F_\infty=
{\rm lim}
\{(\zeta^b\nabla_b\zeta^c) (\zeta^a \nabla_a\zeta_c)/(-\zeta^d\zeta_d)\}
={\rm lim}
\{(\chi^b\nabla_b\chi^c) (\chi^a\nabla_a\chi_c)/(-\chi^d\chi_d)\}
=\kappa_3^2$.
It follows that $\kappa_4$ is invariant under stationary axisymmetric
conformal transformations. More general conformal transformations will
in general destroy the physical interpretation of $\kappa_4$ as the
force per unit mass exerted at infinity to hold a zero angular
momentum particle just outside the horizon.

\section{Hawking radiation and the trace anomaly}

As mentioned in the introduction, the Hawking radiation at infinity in
a conformally coupled free field is invariant under conformal
transformations. But even for a nonconformally coupled field, the
Hawking {\it temperature} (as opposed to the scattering of the
radiation) is conformal invariant. This already follows from conformal
invariance of the surface gravity, together with the known relation
$T_H=\kappa/2\pi$ between surface gravity and Hawking temperature.
However, it is instructive to understand this fact directly in terms
of the derivation of the Hawking effect.

To deduce the Hawking effect, one can compare the free-fall
frequencies of outgoing modes near the horizon with those in the
asymptotic future, rather than by comparing in and out modes. For this
purpose, suppose the line element on a timelike surface has the form
$ds^2=C\, du dv$, where $C$ goes to $0$ at the horizon and to $1$ at
future null infinity, and $u$ goes to infinity as the horizon is
approached. Then $u$ is the retarded Killing time coordinate at future
null infinity, and can be used to define positive frequency there. The
relevant ``free-fall" notion of positive frequency for defining the
Hawking state near the horizon can be taken with respect to the affine
parameter $\lambda$ along a (null) line of constant $v$. This affine
parameter satisfies $d\lambda=C\, du$. As $u\rightarrow\infty$ at the
horizon, $\lambda$ runs over only a finite range. Sufficiently near
the horizon, the effect of a non-singular conformal transformation
$C\rightarrow \wt{C}=\Omega^2 C$ is thus simply to rescale $\lambda$
to $\wt{\lambda}=\Omega_H^2 \lambda$. For the very high
affine-frequency wavepackets near the horizon that are relevant in the
Hawking effect, the notions of positive $\lambda$-frequency and
positive $\wt{\lambda}$-frequency thus agree, so the Hawking
temperature is unchanged.

This argument showing the conformal invariance of the Hawking
temperature for nonconformally coupled fields holds also for
nonstationary conformal transformations. Note however that without
stationarity, the Hawking radiation will be distorted by a
time-dependent ``potential", and there will in general be particle
production over and beyond the Hawking radiation.

The conformal invariance of the Hawking radiation at first seems to
contradict the fact\cite{DavFullUn,ChristFull,BirDav} that the Hawking
energy flux in a conformally coupled field is determined in two
spacetime dimensions by the {\it non}-conformal-invariant trace
anomaly $\langle T\rangle=R/24\pi$ (and is intimately related to the
trace anomaly in {\it any} spacetime dimension). Here we shall briefly
reconcile these two facts.

Assuming the expectation value of the energy-momentum tensor $\langle
T^{ab}\rangle$ is conserved ($\nabla_a \langle T^{ab}\rangle =0$) and
finite, it follows without further assumptions (not even stationarity
\cite{ultrashort}) that the flux at infinity is given in the limit
$u\rightarrow\infty$ by
\begin{equation}
T_{uu}(\infty)=\int_{v_0}^\infty C \langle T\rangle_{,u}\; dv,
\label{fluxint}
\end{equation}
where the metric is $ds^2=C\, du dv$ as in the previous section, and
$v_0$ is any fixed value of $v$. (Actually, $T_{uu}$ signifies the net
outgoing energy flux only if there is no {\it in}coming flux at late
advanced times $v$.) Now $R=2C^{-1}(\ln C)_{uv}$, so we have $C
R_u=2\{(\ln C)_{uu}-\frac{1}{2}[(\ln C)_u]^2\}_v$. Thus in fact the
above formula (\ref{fluxint}) for $T_{uu}$ can be integrated quite
generally, yielding
\begin{equation}
T_{uu}(\infty)=\frac{-1}{12\pi}
[(\ln C)_{uu}-\frac{1}{2}(\ln C)_u^2] (u=\infty, v_0).
\label{flux}
\end{equation}
Under a regular conformal transformation $C\rightarrow\wt{C}=\Omega^2 C$,
the right hand side of (\ref{flux}) only changes by
$u$-derivatives of $\ln \Omega^2$. Regularity of $\Omega$ implies that
$\Omega_u=(d\Omega/d\lambda)(d\lambda/du)$ vanishes at the horizon,
because $d\Omega/d\lambda$ is finite there and $d\lambda/d u$ vanishes
since an infinite range of the $u$ coordinate is covered in a finite
range of affine parameter $\lambda$. Thus the flux $T_{uu}(\infty)$ is
in fact unchanged by a regular conformal transformation, as expected.

\vskip 3em
It is a pleasure to thank David Garfinkle and Ulvi Yurtsever for a
helpful discussion on conformal Killing fields and Killing horizons,
and D.G. for a correction and suggestions that improved the
presentation. T.J.~was supported by NSF grant PHY91-12240. Research at
the ITP, UCSB was supported by NSF Grant PHY89-04035.

\end{document}